\documentclass[]{pkuai4m}

\usepackage[toc,page,header]{appendix}

\usepackage{minitoc}
\usepackage{multirow}     
\usepackage{multicol}     
\usepackage{booktabs}     
\usepackage{colortbl}     
\usepackage[table]{xcolor} 
\usepackage{makecell}     
\usepackage{graphicx}     
\usepackage{minted}
\usepackage{xcolor}
\usepackage{fontawesome5}

\title{Matlas: A Semantic Search Engine for Mathematics}

\author[1,3*]{Haocheng Ju}
\author[2,3,7*]{Leheng Chen}
\author[3]{Peihao Wu}
\author[3]{Bryan Dai}
\author[4,5,6,7\ddagger]{Bin Dong}

\affiliation[]{
$^{1}$School of Mathematical Sciences, Peking University, Beijing 100871, China\\
$^{2}$Beijing International Center for Mathematical Research, Peking University, Beijing 100871, China \\
$^{3}$IQuest Research \\
$^{4}$Beijing International Center for Mathematical Research and the New Cornerstone Science Laboratory, Peking University, Beijing 100871, China \\
$^{5}$Center for Machine Learning Research, Peking University, Beijing 100871, China \\
$^{6}$Center for Intelligent Computing, Great Bay Institute for Advanced Study, Great Bay University, Dongguan 523000, China \\
$^{7}$Zhongguancun Academy, Beijing 100094, China \\
}

\contribution[*]{Equal Contribution}
\contribution[\dagger]{Corresponding author}

\abstract{
Retrieving mathematical knowledge is a central task in both human-driven research, such as determining whether a result already exists, finding related results, and identifying historical origins, and in emerging AI systems for mathematics, where reliable grounding is essential. However, the scale and structure of the mathematical literature pose significant challenges: results are distributed across millions of documents, and individual statements are often difficult to interpret in isolation due to their dependence on prior definitions and theorems. In this paper, we introduce Matlas, a semantic search engine for mathematical statements. Matlas is built on a large-scale corpus of 8.07 million statements extracted from 435K peer-reviewed papers spanning 1826 to 2025, drawn from a curated set of 180 journals selected using an ICM citation-based criterion, together with 1.9K textbooks. From these sources, we extract mathematical statements together with their dependencies, construct document-level dependency graphs, and recursively unfold statements in topological order to produce more self-contained representations. On top of this corpus, we develop a semantic retrieval system that enables efficient search for mathematical results using natural language queries. We hope that Matlas can improve the efficiency of theorem retrieval for mathematicians and provide a structured source of grounding for AI systems tackling research-level mathematical problems, and serve as part of the infrastructure for mathematical knowledge retrieval.
}

\date{\today}

\def\emailicon{\raisebox{-1.5pt}{\includegraphics[height=1.05em]{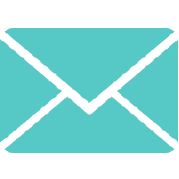}}}

\def\platformicon{\raisebox{-1.5pt}{\includegraphics[height=1.05em]{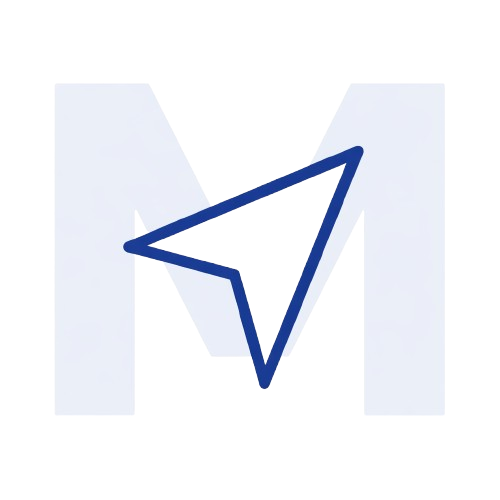}}}

\checkdata[ \emailicon \hspace{0.3em} Correspondence ]{\email{dongbin@math.pku.edu.cn}}

 \newcommand{\sourcelink}{https://github.com/frenzymath}
 \newcommand{\datalink}{https://huggingface.co/FrenzyMath/matlas}
 \newcommand{\platformlink}{https://matlas.ai/}
 \newcommand{\apidocslink}{https://matlas.ai/docs}

\checkdata[ \platformicon \hspace{0.3em} Web Service ]{ \url{\platformlink} }
\checkdata[ \platformicon \hspace{0.3em} API Docs ]{ \url{\apidocslink} }

\begin{document}
\maketitle

\renewcommand{\thefootnote}{\fnsymbol{footnote}} 
\setcounter{footnote}{0}

\renewcommand{\thefootnote}{\arabic{footnote}}
\pagestyle{fancy}
\fancyhf{}
\fancyhead[R]{\thepage}


\section{Introduction}
\begin{figure}[tb]
\centering
\includegraphics[width=0.6\linewidth]{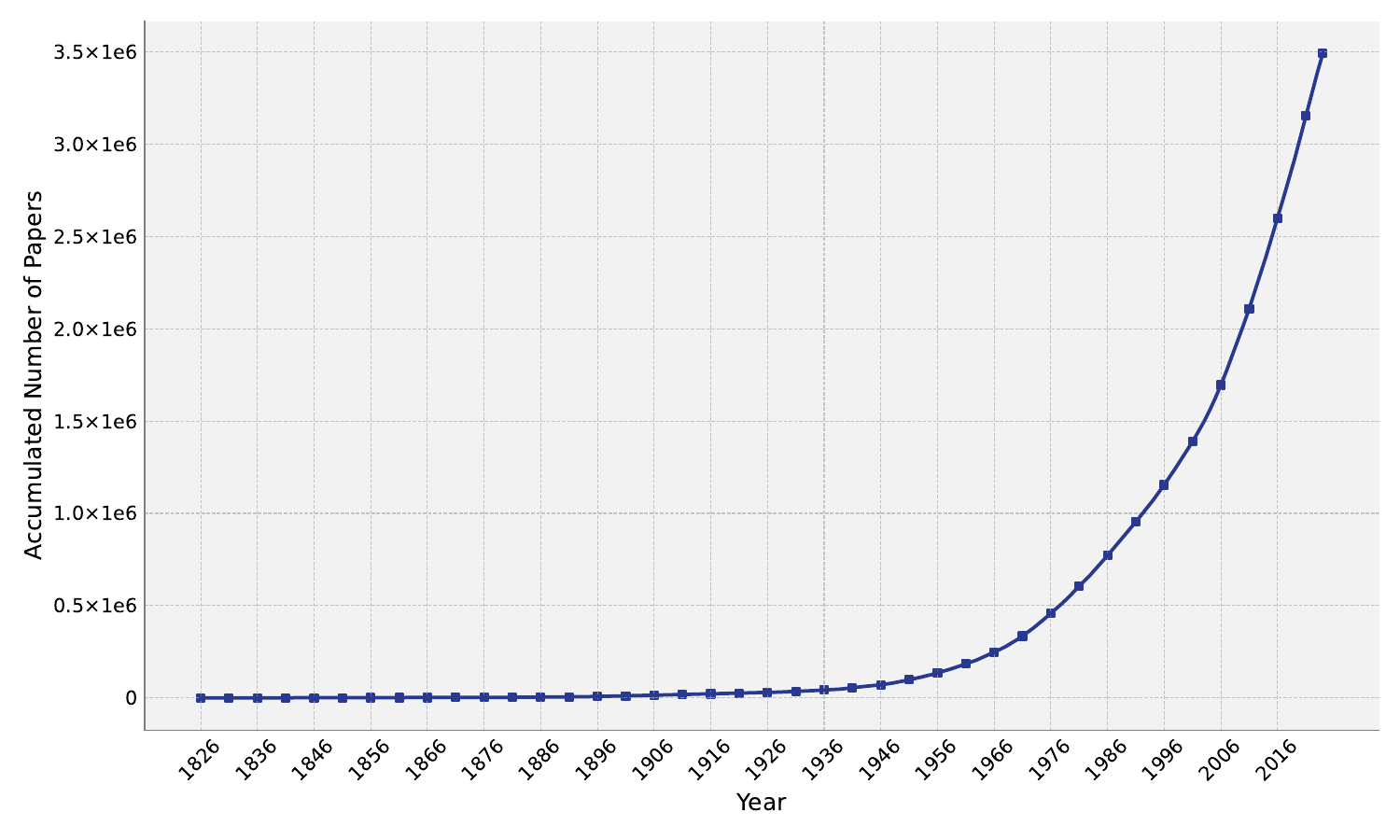}
\caption{The cumulative number of mathematics journal papers from 1826 to 2024 (across all journals, not restricted to our selected set of 180), computed from MathSciNet metadata by counting entries with entryType = "J" as journal publications.}
\label{fig:paper_count}
\end{figure}

Retrieving mathematical results is a fundamental activity in mathematical research. Discovering related results and variants is essential for advancing new work, while determining whether a result already exists helps avoid redundant effort and overlooked prior contributions. Identifying the original source and historical origin of a theorem is also crucial for proper attribution. However, the mathematical literature has grown dramatically over time (Figure \ref{fig:paper_count}) and now contains millions of articles and books, making it increasingly challenging to locate relevant results efficiently.

Beyond its importance for human researchers, mathematical retrieval is also a key component for AI systems aiming to perform research-level mathematics. Recent advances in large language models (LLMs) have enabled them to tackle certain research-level problems \cite{feng2026aletheia,feng2026eigenweights,feng2026towards,patel2026simplicity,lee2026lower}. However, grounding such systems in existing mathematical knowledge is likely to be beneficial. In particular, two aspects are critical: reliability and self-containment. For reliability, peer-reviewed publications are generally more trustworthy than arXiv preprints, despite the possibility of errors. For self-containment, mathematical statements are inherently dependent on prior definitions and results; as a consequence, an isolated statement is often difficult to interpret without its dependencies. This motivates the need to recursively unfold statements so that they become self-contained and interpretable.

Motivated by these considerations, namely supporting human mathematicians and providing stronger grounding for AI systems, we develop \textit{Matlas}, a semantic search engine for mathematics. Matlas is built on a large-scale corpus comprising 8.07 million mathematical statements extracted from 435K published papers and 1.9K textbooks. The papers are drawn from a curated set of 180 journals selected via an ICM citation-based criterion, spanning the years 1826 to 2025. From these sources, we extract statements together with their dependencies, construct directed dependency graphs for each document, and unfold statements in topological order to obtain more self-contained representations. On top of this corpus, we build a semantic search system to enable efficient retrieval of mathematical results. 

An early version of Matlas was used in a case study where it helped identify a key technical result and supported a natural language reasoning agent in resolving an open conjecture autonomously \cite{ju2026automated}, suggesting its potential value in mathematical research.

\section{Related Works}
In this section, we review prior work on semantic search engines for both formal and natural language mathematics. For Lean’s mathematical library mathlib4, several semantic search systems have been developed, including Moogle\footnote{\href{https://www.moogle.ai/}{https://www.moogle.ai/}}
, LeanSearch \cite{gao-etal-2024-semantic-search}, LeanExplore \cite{asher2025leanexplore}, LeanFinder \cite{lu2025lean}, and LeanDex \footnote{\href{https://leandex.projectnumina.ai/}{https://leandex.projectnumina.ai/}}
. Most of these systems construct an aligned informal–formal corpus derived from mathlib4 and perform retrieval over either the informal corpus alone or a combination of informal and formal representations. For natural language mathematics, \cite{alexander2026semantic} constructs a semantic search engine for mathematical statements using a corpus drawn primarily from arXiv papers, with a smaller portion from sources such as ProofWiki, the Stacks Project, and the Open Logic Project. For arXiv papers, they extract theorem statements from LaTeX sources and use an LLM to generate a short natural language description for each statement based on the statement itself and the introduction of the paper. This description serves as the representation of the theorem and is used for embedding, rather than the original statement. Our work differs from \cite{alexander2026semantic} in two main aspects. First, our corpus consists of published papers from a curated set of 180 journals (spanning 1826 to 2025) together with textbooks, rather than arXiv papers, whose coverage begins in 1991. Second, we aim to make statements self-contained: we extract dependencies between statements, construct a directed graph for each document, and unfold the statements hierarchically based on this dependency structure.

\section{Methodology}
\begin{figure}[tb]
\centering
\includegraphics[width=0.99\linewidth]{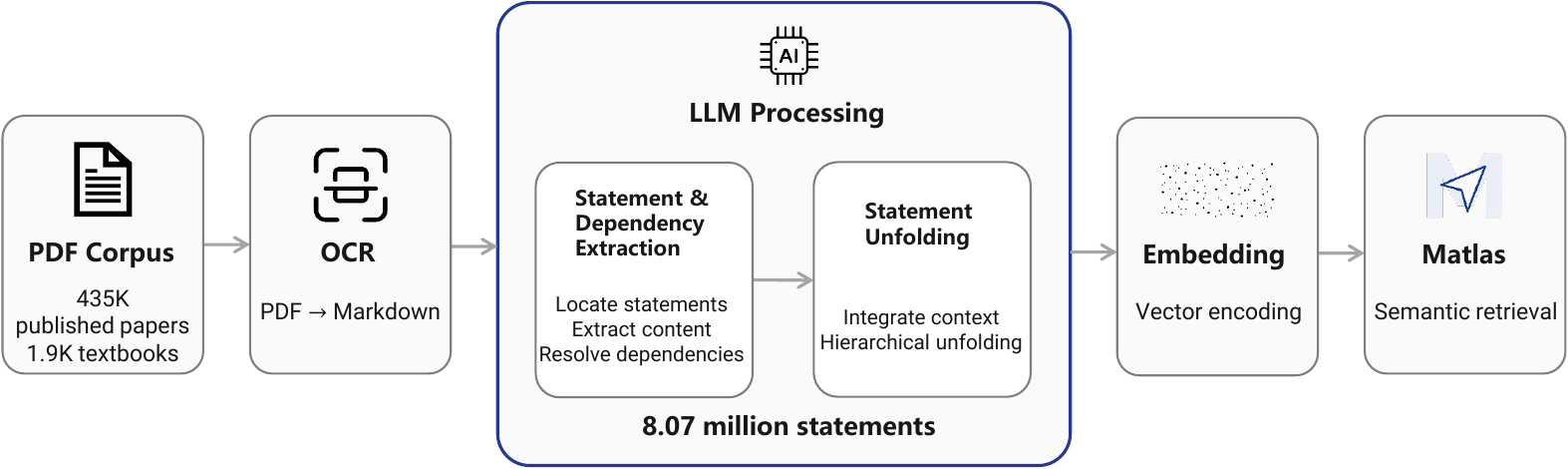}
\caption{Overview of our pipeline.}
\label{fig:overview}
\end{figure}

In this section, we describe our methodology for building Matlas. Figure~\ref{fig:overview} provides an overview of the framework. To construct a theorem search engine based on relatively reliable sources, we select a set of 180 journals using an ICM citation-based criterion and collect 435K published papers spanning 1826 to 2025 from these journals, together with 1.9K textbooks. While peer review does not guarantee absolute correctness, it generally provides a higher level of reliability than preprints such as those on arXiv. We then apply OCR to the PDFs to obtain raw text.

From this corpus, we extract 8.07 million mathematical statements together with their dependencies, since many statements rely on preceding definitions or results. For each paper or book, we construct a directed graph capturing these dependencies, and then unfold the statements in topological order to produce self-contained statements.

Next, we build the semantic search engine by embedding the processed statements into vector representations and indexing them in a vector database. Given a query, we embed it into a vector and retrieve relevant statements using cosine similarity as the ranking metric. We begin by describing the corpus selection process in Section~\ref{subsec:corpus}, followed by statement and dependency extraction in Section~\ref{subsec:extract}, and finally the statement unfolding procedure in Section~\ref{subsec:unfold}.

\subsection{Corpus Construction}\label{subsec:corpus}
We construct a set of 180 journals through an ICM-based filtering process. Specifically, we first gather the full collection of ICM proceedings and extract the references cited in each paper. For every journal, we then count how many times its papers published between 2007 and 2021 are cited by these proceedings. We keep only those journals that receive more than 50 such citations and that have published at least 100 papers over the same period. We obtain metadata for papers in these journals from MathSciNet and collect 606K PDFs of such papers, which represent a subset of all papers published in these journals throughout their history. Later, we find that only 435K of them contain mathematical statements. Some papers indexed by MathSciNet, such as certain applied mathematics, computational physics, and computational chemistry papers, do not contain mathematical statements. For textbooks, we compile a book list from five publishers: Springer, Princeton, Cambridge, Oxford, and AMS. This list contains 3.3K textbooks, of which we are able to collect 1.9K. After collecting the papers and textbooks, we apply OCR to the PDFs to prepare them for further processing.

\subsection{Statement and Dependency Extraction}\label{subsec:extract}

In this subsection, we focus on extracting statement-level units from mathematical texts, such as definitions, theorems, and lemmas, together with their dependency relations. A straightforward approach would be to either provide the entire document to a large language model or divide the document into smaller chunks and process them independently. However, both approaches are not well-suited for mathematical content: the full document often exceeds model context limits, while naive chunking can disrupt the internal structure of statements by splitting them across boundaries.

\begin{figure}[tb]
\centering
\includegraphics[width=0.98\linewidth]{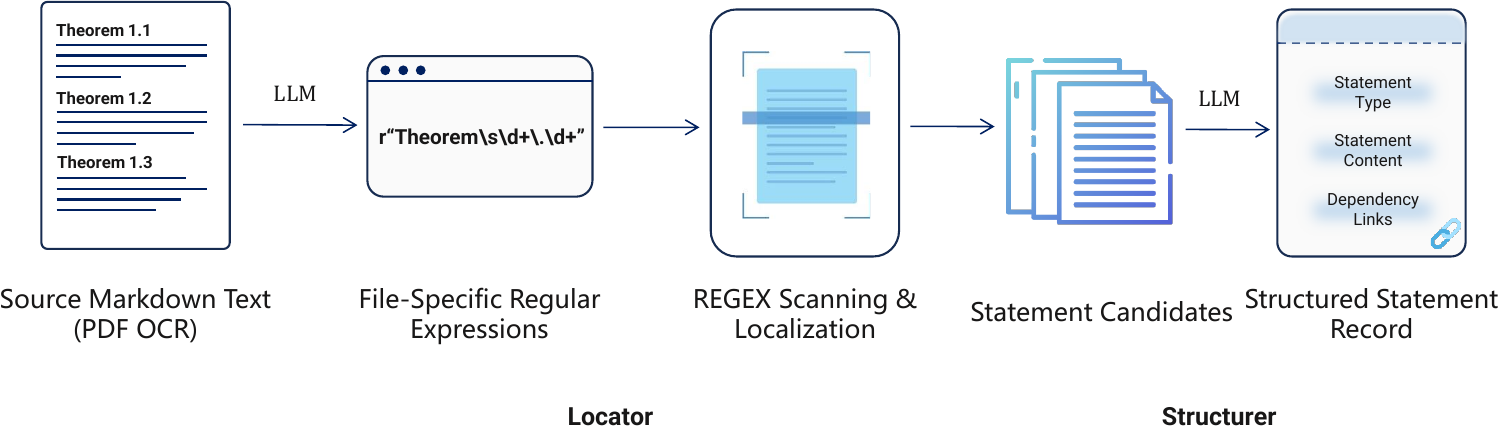}
\caption{Two-stage pipeline for statement and dependency extraction. The locator first identifies text segments corresponding to individual statement units. The structurer then transforms each localized span into a structured representation, including the statement type, content, and local dependency links.}
\label{fig:statement-extraction-pipeline}
\end{figure}

To overcome these issues, we design a two-stage pipeline (Figure~\ref{fig:statement-extraction-pipeline}). In the first stage, an LLM\footnote{We use DeepSeek-V3.2 \cite{liu2025deepseek} as the LLM.} processes the markdown text produced by PDF OCR and generates document-specific regular expressions that reflect local formatting patterns of statements. These patterns are then applied to the text to identify candidate statement spans, each corresponding to a single statement unit. The objective of this stage is to reliably localize such spans and produce stable, statement-centered inputs for subsequent processing.

In the second stage, the extracted candidates are handled in small, partially overlapping batches, with a default batch size of five. For each batch, we construct forward context windows around the candidate positions in the OCR markdown text, using a default window length of 4000 characters. Overlapping windows are merged before being passed to the model. Based on this merged context, the LLM structurer transforms each candidate into a structured representation that includes its type, textual content, and local dependency links. The use of overlapping sliding windows helps preserve short-range dependencies that might otherwise be lost at batch boundaries. In this way, structured extraction is performed on already localized statement units.

Extracting dependencies is a key component of the pipeline. Mathematical statements are often not self-contained, as they depend on earlier definitions, notation, or intermediate results. Therefore, the system also identifies local dependency relationships among statements. This information is valuable both for understanding the structure of the text and for supporting the subsequent unfolding stage.

\subsection{Statement Unfolding}\label{subsec:unfold}

\begin{figure}[tb]
\centering
\includegraphics[width=0.55\linewidth]{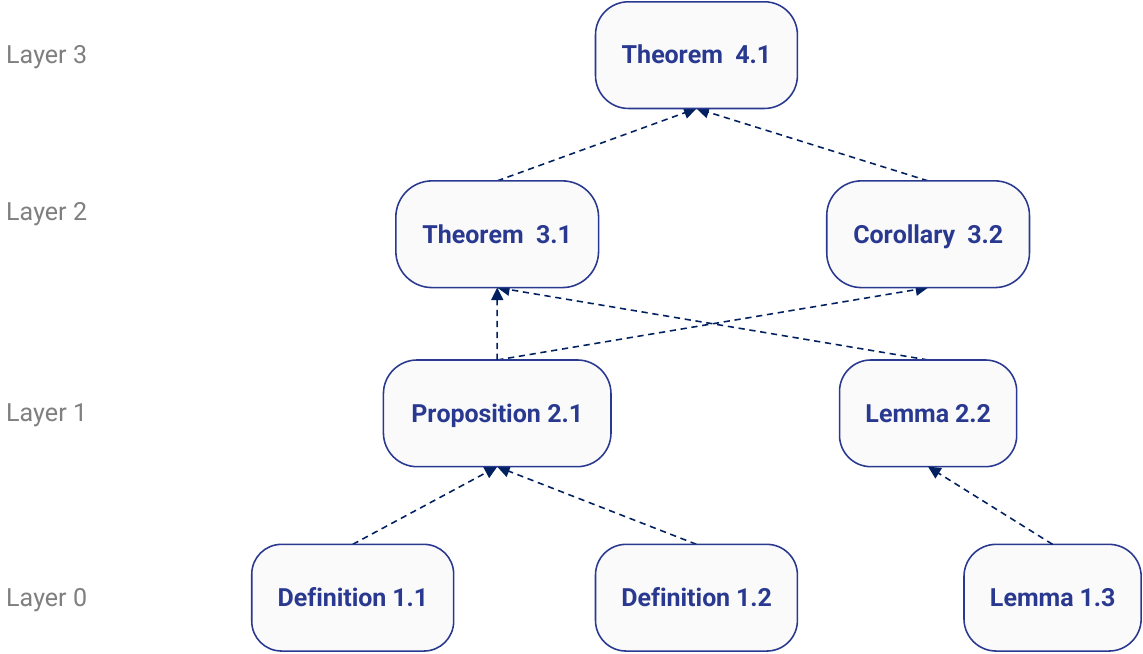}
\caption{An example of the directed dependency graph and its layered structure. Statement unfolding proceeds from layer 0 to the highest layer.}
\label{fig:dag}
\end{figure}

After extracting statements and their dependencies from each paper and textbook, we construct a directed graph for each document. Each node represents a mathematical statement, and there is an edge from node A to node B if and only if B depends on A. An example is shown in Figure \ref{fig:dag}. To make statements self-contained, we then unfold this graph in topological order. Specifically, we first partition the graph into layers, where each layer consists of nodes with zero in-degree after removing all nodes from the preceding layers. We then process the layers sequentially, starting from layer 0 to the final layer, recursively expanding each statement using the already processed statements from earlier layers.

We note that \cite{peyronnet2026lemmabench} also aims to make statements in mathematical papers self-contained by extracting the definitions and assumptions required for a given statement. They explore two approaches. The first is full-context retrieval, where all content preceding the target statement is provided to an LLM, which then identifies the relevant definitions and assumptions. The second is vector-based retrieval, where an LLM first extracts non-trivial objects appearing in the statement; regular expressions are then used to locate paragraphs containing these objects, followed by dense retrieval to select the top-$k$ candidate paragraphs. These paragraphs, together with the target objects, are then fed into an LLM to extract the necessary definitions and assumptions. Our method differs from theirs in that we do not attempt to recover all dependent statements in a single pass. Instead, we extract only first-order dependencies (corresponding to the previous layer) and expand statements incrementally in a layer-by-layer manner.

After obtaining the unfolded statements, totaling 8.07 million, we build our theorem search engine using dense retrieval. Specifically, we encode each unfolded statement into a vector representation using Qwen3-Embedding-8B \cite{zhang2025qwen3} and store these vectors in a vector database. Given a query, we prepend a theorem-retrieval-oriented instruction to the input, encode the resulting text into a vector, and retrieve relevant statements from the database using cosine similarity.

\section{Conclusion}\label{sec: concl}
In this paper, we develop Matlas, a semantic search engine for mathematical statements drawn from relatively reliable sources. The underlying corpus is constructed by collecting PDFs of papers published in a curated set of 180 journals, selected using an ICM citation-based criterion to filter out relatively low-quality venues, together with 1.9K textbooks. These documents are processed through two phases: statement and dependency extraction, and statement unfolding. In the statement and dependency extraction phase, we first identify the positions of individual statements, and then assemble batches of statements along with their surrounding context to input to an LLM for dependency extraction. In the statement unfolding phase, we construct a directed graph for each document based on the extracted dependencies, and unfold statements in topological order to obtain more self-contained representations.

We hope that Matlas can improve the efficiency of theorem retrieval for mathematicians and provide stronger grounding for AI systems tackling research-level mathematical problems using reliable sources. By organizing mathematical knowledge at the level of individual statements and their dependencies, Matlas may also facilitate the transfer and reuse of tools and results across different areas of mathematics. Overall, we view Matlas as a step toward building infrastructure for mathematical knowledge retrieval that supports both human and AI-driven mathematical discovery.

\section*{Acknowledgement}
We thank DP Technology for providing the PDFs of the published papers and for performing OCR processing. This work is supported in part by the National Key R\&D Program of China grant 2024YFA1014000, the Fundamental and Interdisciplinary Disciplines Breakthrough Plan of the Ministry of Education of China (JYB2025XDXM113), and the New Cornerstone Investigator Program.

\clearpage

\bibliographystyle{plainnat}
\bibliography{ref}



\end{document}